\documentclass[%
 reprint,%
 amssymb, amsmath,%
 aip,cha,%
]{revtex4-2}
\usepackage{graphicx}
\usepackage{amsmath}
\usepackage{dcolumn}
\usepackage{bm}

\usepackage{multirow}
\usepackage{diagbox}
\usepackage{bm}%
\expandafter\ifx\csname package@font\endcsname\relax\else
 \expandafter\expandafter
 \expandafter\usepackage
 \expandafter\expandafter
 \expandafter{\csname package@font\endcsname}%
\fi
\begin{document}

\title[]{Quantum Emission in Monolayer WSe$_2$ Transferred onto InP Nanowires}
\author{Palwinder Singh}
\thanks{These authors contributed equally.}
\affiliation{Department of Physics and Atmospheric Science, Dalhousie University, Halifax, Nova Scotia, B3H 4R2, Canada}
\affiliation{National Research Council Canada, Ottawa, Ontario, K1A 0R6, Canada}
\author{Jasleen Kaur Jagde}
\thanks{These authors contributed equally.}
\affiliation{Department of Physics and Atmospheric Science, Dalhousie University, Halifax, Nova Scotia, B3H 4R2, Canada}
\author{Megha Jain}
\affiliation{National Research Council Canada, Ottawa, Ontario, K1A 0R6, Canada}
\affiliation{Centre for Nanophotonics, Department of Physics, Engineering Physics, and Astronomy, Queen's University, Kingston, Ontario, K7L 3N6, Canada}
\author{Edith Yeung}
\affiliation{National Research Council Canada, Ottawa, Ontario, K1A 0R6, Canada}
\affiliation{Department of Physics, University of Ottawa, Ottawa, Ontario, K1N 6N5, Canada}
\author{David B. Northeast}
\affiliation{National Research Council Canada, Ottawa, Ontario, K1A 0R6, Canada}
\author{Simona Moisa}
\affiliation{National Research Council Canada, Ottawa, Ontario, K1A 0R6, Canada}
\author{Seid J. Mohammed}
\affiliation{National Research Council Canada, Ottawa, Ontario, K1A 0R6, Canada}
\author{Jean Lapointe}
\affiliation{National Research Council Canada, Ottawa, Ontario, K1A 0R6, Canada}
\author{Una Rajnis}
\affiliation{Department of Physics and Atmospheric Science, Dalhousie University, Halifax, Nova Scotia, B3H 4R2, Canada}
\author{Annika Kienast}
\affiliation{Department of Physics and Atmospheric Science, Dalhousie University, Halifax, Nova Scotia, B3H 4R2, Canada}
\author{Philip J. Poole}
\affiliation{National Research Council Canada, Ottawa, Ontario, K1A 0R6, Canada}
\author{Dan Dalacu}
\email{dan.Dalacu@nrc-cnrc.gc.ca}
\affiliation{National Research Council Canada, Ottawa, Ontario, K1A 0R6, Canada}
\affiliation{Centre for Nanophotonics, Department of Physics, Engineering Physics, and Astronomy, Queen's University, Kingston, Ontario, K7L 3N6, Canada}
\affiliation{Department of Physics, University of Ottawa, Ottawa, Ontario, K1N 6N5, Canada}
\author{Kimberley C. Hall}
\email{Kimberley.Hall@dal.ca}
\affiliation{Department of Physics and Atmospheric Science, Dalhousie University, Halifax, Nova Scotia, B3H 4R2, Canada}

\date{\today}

\begin{abstract}
Localized quantum emitters in transition-metal dichalcogenides (TMDs) have recently emerged as solid-state candidates for on-demand sources of single photons.  Due to the role of strain in the site-selective creation of TMD emitters, their hybrid integration into photonic structures such as cavities and waveguides is possible using pick-and-place methods. Here we investigate quantum emission from a hybrid structure consisting of a monolayer of WSe$_2$ interfaced with horizontally aligned InP nanowires (NWs).  Our experiments reveal multiple narrow and bright emission peaks in the 715-785 nm spectral range and $g^{(2)}(0)$ as low as 0.049, indicating strong antibunching and good single photon purity.  The faceted nature of III-V NWs provides unique opportunities for strain engineering, including the potential for placement of emitters on the top surface for optimal coupling. Our findings pave the way for realizing hybrid quantum light sources for integrated quantum photonics that could combine III-V quantum dots with TMD emitters into a single platform.
\end{abstract}

\maketitle

\section{Introduction}
The development of high-performance, on-demand solid-state quantum emitters (QEs) would enable the replacement of probablistic single photon sources for applications such as quantum communication \cite{Wehner:2018,Zhang:2025} and photonic quantum computing \cite{Kok:2007,Pan:2012}. Localized QEs in transition-metal dichalcogenides (TMDs) such as WSe$_2$ have shown considerable promise in recent years for application to single photon sources \cite{yu2021site,zhao2021site, lee2024ferroelectric,singh2025exploring}.  These emitters are produced by inducing strain in a monolayer flake, which leads to hybridization of dark excitons with Se vacancy defect states and associated strong radiative recombination \cite{Lindhart:2019}.  Since the discovery of QEs in monolayer WSe$_2$ \cite{srivastava2015optically,he2015single,koperski2015single, chakraborty2015voltage,tonndorf2015single,branny2017deterministic,palacios2017large}, emitters have also been observed in MoSe$_2$ \cite{Branny:2016,yu2021site}, MoS$_2$ \cite{Klein:2021,Barthelmi:2020}, and MoTe$_2$ \cite{zhao2021site}, with a range of emission wavelengths including the telecommunication bands.  Controlling the direction and magnitude of the local strain on the monolayer by varying the substrate and/or using piezoelectric devices enables the optimization of optical properties such as the emission wavelength, polarization and brightness \cite{Sortino:2021,Cai:2024,Kumar:2016,Drawer:2023,iff:2019dynamicstrain,so:2021polarization}.  Single photon emitter rates as high as 69 MHz \cite{Sortino:2021,Drawer:2023} and good single photon purity (g$^2$(0) < 0.002 \cite{Kumar:2016}) have been observed in recent years.  TMD QEs have also been used to demonstrate the BB84 protocol in quantum key distribution \cite{Gao:2023}, showing competitive performance relative to III-V semiconductor quantum dots and NV Centers.  This progress highlights the potential of strain-induced TMD QEs for integrated quantum photonic applications. 
 
 Functional quantum photonic technologies require not only the generation of QEs, but also their deterministic coupling with photonic circuits to route, manipulate, and detect single photons \cite{so2025deterministic, rajsingle}. Hybrid integration approaches, including pick-and-place methods, facilitate the integration of disparate functionalities onto the same chip since the quantum emitter and other components can be independently optimized \cite{Sartison:2022}. This integration presents a dual challenge: the emitter must be positioned at a predefined location, and its emission characteristics must be spectrally matched to the optical modes of the photonic structure \cite{so2025deterministic, rajsingle}.  The strain-induced nature of TMD quantum emitters offers a significant advantage since local strain centers may be used to induce QEs with optimized positions in cavites and waveguides.   The monolayer thickness of TMD materials places the emitters directly on the surface, avoiding photon extraction challenges tied to embedded emitters in high-index materials and facilitating efficient coupling to both on-chip and off-chip photonic components \cite{so2025deterministic}.  This approach has been used to successfully integrate TMD QEs into optical cavities using Purcell enhancement to achieve directional emission control and higher photon rates \cite{Luo:2018,Cai:2024,iff:2018,Fryett:2018,iff2021purcell,Drawer:2023}, and in both metallic and dielectric waveguides for photon routing \cite{blauth2018coupling,Dutta:2018,Cai:2017,ErrandoHerranz:2021,peyskens2019integration,Tonndorf:2017,White:2019,Kim:2019,Schell:2017}.     Precise control over the geometry and magnitude of the strain is essential for deterministic and scalable integration of quantum emitters into these photonic systems \cite{ErrandoHerranz:2021}. 
 
 Nanowire (NW) waveguides containing single InAsP quantum dots offer excellent performance as single photon sources, exhibiting ultra-narrow emission linewidths \cite{Laferriere2023approaching}, collection efficiencies above 80\% \cite{laferriere2022unity} and g$^2$(0) below 0.01 \cite{laferriere2022unity}.  The successful integration of these NW single photon sources with silicon nitride waveguides has also been demonstrated in recent years \cite{Yeung:2023}.  The integration of such NW sources with TMD emitters, especially telecom-compatible systems \cite{zhao2021site}, would permit multiplexing of single photon sources and open up possibilities for generating coupled emitters for wavelength transduction.  In this work, we demonstrate the formation of strain-induced quantum emitters in monolayer WSe$_2$ deposited onto InP NWs. We observe bright and narrow emission lines in photoluminescence (PL) spectroscopy from a monolayer WSe$_2$/InP NW hybrid structure. The g$^2$(0) values range from 0.049 to 0.24, consistent with strong antibunching. Characterization of local strain using atomic force microscopy (AFM) highlights the unique role of the NW faceted surfaces in strain engineering. Our findings lay the foundation for combining TMD emitters with III-V semiconductor integrated optics.

\section{Experimental Section}
InP NWs were grown on InP substrates via selective-area vapor-liquid-solid epitaxy \cite{dalacu2009selective}. The NWs have a base diameter of 250\,nm, tapering to a few nanometers at the tip and extending to several micrometers in length. Detailed growth information can be found elsewhere \cite{dalacu2009selective, dalacu2021tailoring}. These NWs were picked up and transferred onto a Si/SiO$ _2 $ substrate using polydimethylsiloxane (PDMS, GEL PAK 4). Thin layers of WSe$ _2 $ (HQ Graphene) were obtained through mechanical exfoliation using adhesive tape \cite{Gomez:2014}. These thin layers were picked up using PDMS. Monolayers were identified under an optical microscope (Nikon Eclipse ME600) and transferred to a Si/SiO$ _2 $ substrate with NWs. This transfer was performed using a transfer station equipped with a microscope to allow precise placement of the monolayer over the targeted area. 

\begin{figure}[b]
\includegraphics[width=9.0cm]{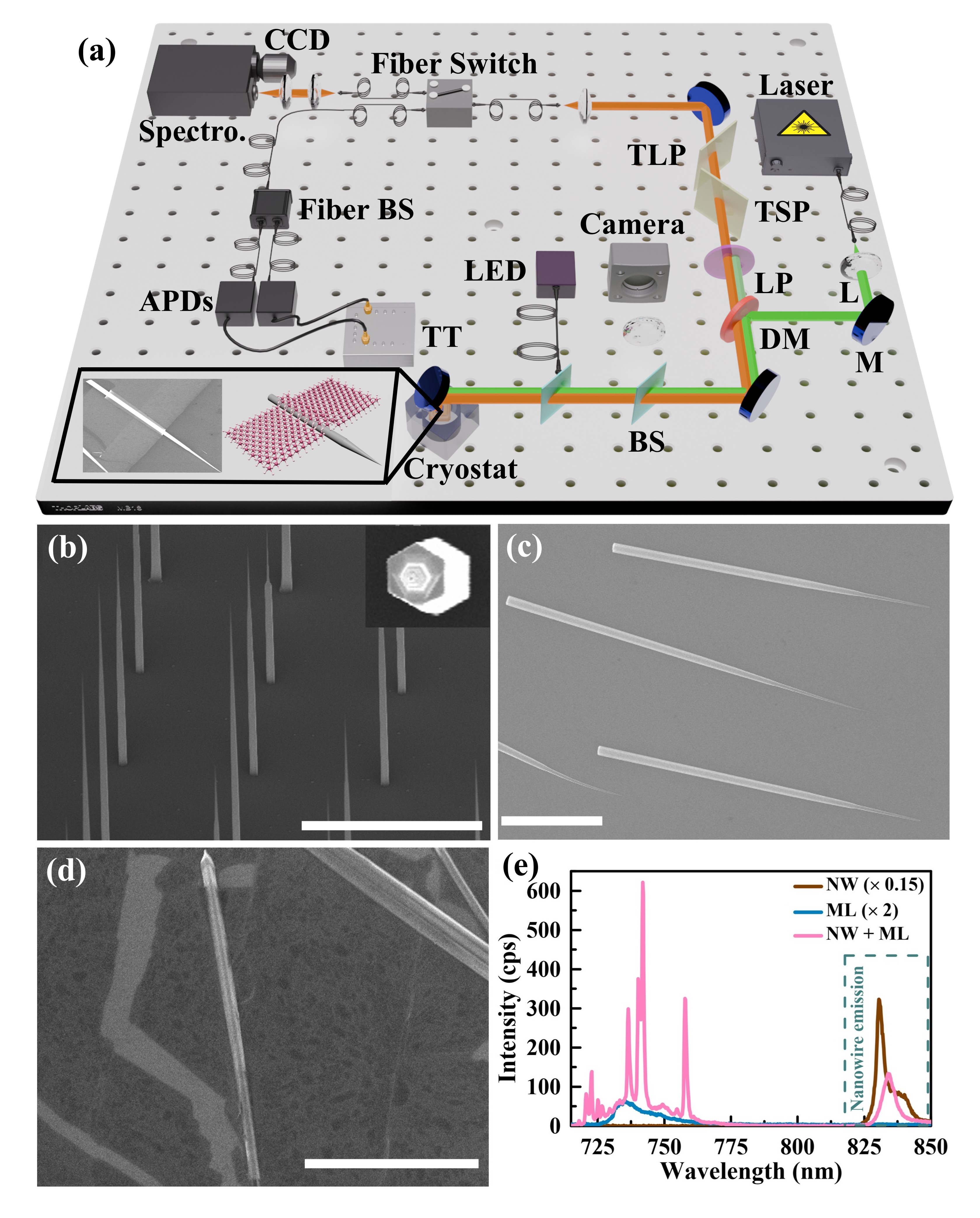}
\caption{(a) Schematic diagram of the optical setup; L: lens, M: mirror, DM: dichroic mirror, BS: beam splitter, LP: long pass filter, TLP: tunable long pass filter, TSP: tunable short pass filter, TT: time tagger, APD: avalanche photo-diode, CCD: charge coupled device.  (b) SEM image of an as-grown array of InP NWs at an angle of 45$^\circ$.  Inset shows a top view image of a NW with a base diameter of 250\,nm. (c) SEM image of InP NWs horizontally deposited onto an Si/SiO$ _2 $ substrate. (d) Monolayer WSe$ _2 $/NW hybrid structure. (e) PL spectra of a bare NW, an unstrained monolayer, and a monolayer on NW with excitation powers of 10 $  \mu$W, 15 $\mu$W, and 10 $  \mu$W, respectively. Scale bars: 10 $  \mu$m in (b), 5 $\mu$m in (c), and 3 $\mu$m in (d).
}
\label{Figure_1}
\end{figure}

PL spectroscopy was conducted at 4 K using a $\mu$-PL setup and a 520\,nm continuous wave laser as an excitation source. A schematic diagram of the optical setup is shown in Fig.~\ref{Figure_1}(a). PL emission from the monolayer was collected using an aspheric cryogenic objective (numerical aperture: 0.55) and coupled to a single-mode fiber. The collected light was routed to a spectrometer (Princeton Instruments) fitted with 150 g/mm and 1200 g/mm gratings, and detected using a CCD camera (Pixis-100BR eXcelon). Second-order autocorrelation measurements were performed using a fiber-coupled Hanbury Brown and Twiss setup. The emission from strained monolayer was filtered through a combination of sharp edge long-pass (TLP01-790, Semrock) and short-pass (TSP01-790, Semrock) filters before being coupled into a single mode fiber. The filtered emission was then directed to two fiber-coupled silicon single photon avalanche photodiodes via a 50:50 fiber-coupled beam splitter. Scanning electron microscopy (SEM) was performed on as-grown InP NWs, NWs transferred onto an Si/SiO$ _2 $ substrate, and monolayer WSe$ _2 $ transferred onto NWs using a Hitachi FESEM (S-4700). AFM measurements were carried out using a Bruker-Veeco Nanoscope IIIa in tapping mode, with a TESPA-V2 probe (spring constant: k = 42\,Nm$^{-1}$).

\section{Results and Discussion}

\begin{figure}[h]
\centering
\includegraphics[scale=0.35]{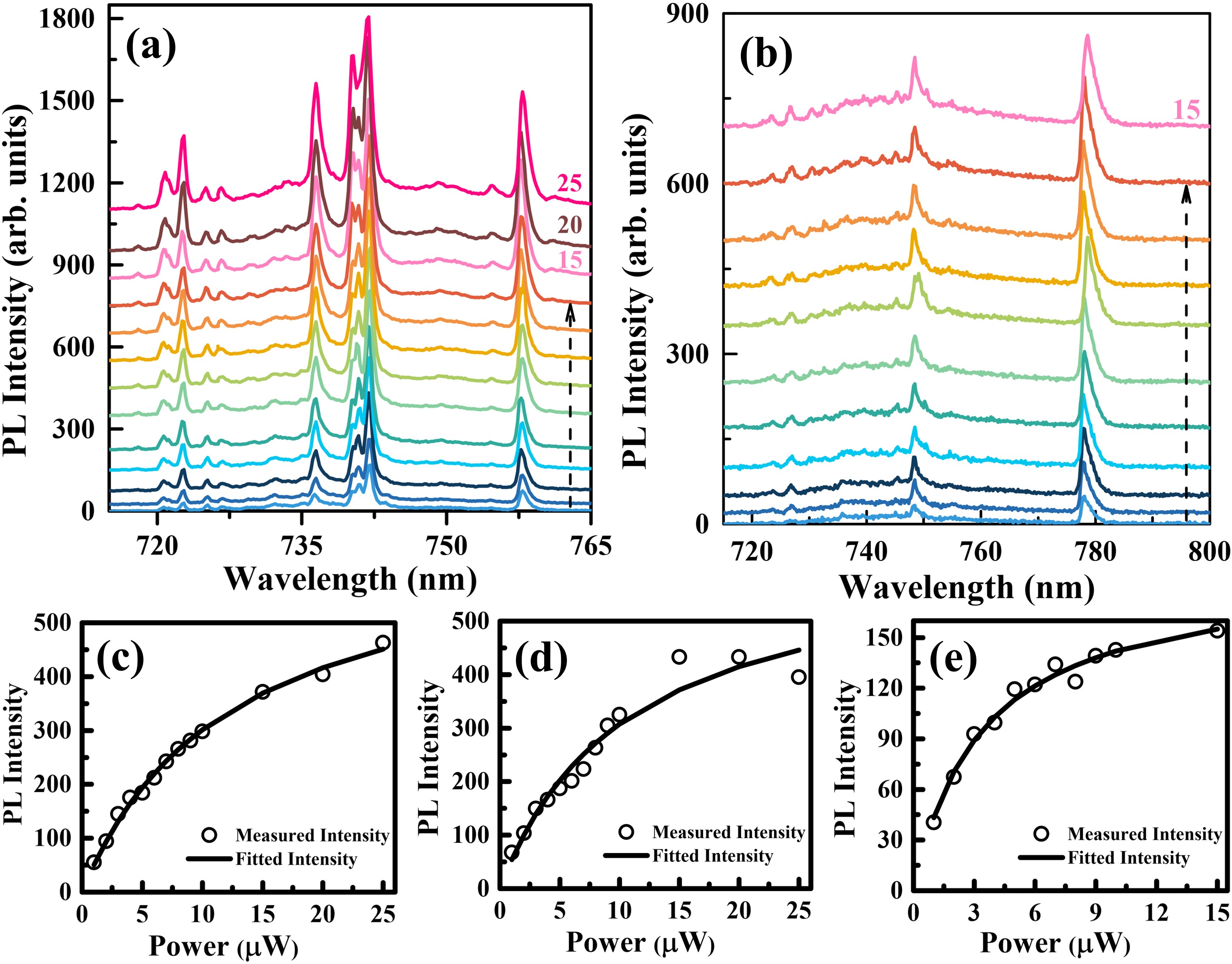}
\caption{PL spectra collected at various excitation powers from strained spots S2 [(a)] and S3 [(b)]. The excitation power increases from 1\,$\mu$W to 10\,$\mu$W in the direction of the arrow, with other power values labeled directly on the respective spectra. PL intensity dependence on excitation power for quantum emissions at (c) 736.47\,nm from strained spot S2, (d) 757.70\,nm from strained spot S2, and (e) 778.37\,nm from strained spot S3. The solid line represents a fit based on the saturation model.}
\label{Figure_3}
\end{figure}

Fig.~\ref{Figure_1}(b) displays an SEM image of vertically aligned InP NWs epitaxially grown on an InP substrate. Fig.~\ref{Figure_1}(c) shows an SEM image of NWs after mechanical transfer onto an Si/SiO$_2$  substrate, forming a horizontally aligned random ensemble that serves as a platform for strain engineering. Fig.~\ref{Figure_1}(d) presents an SEM image of a monolayer of WSe$_2$ transferred onto a long NW, illustrating the formation of a hybrid structure in which the monolayer conforms to the underlying NW topography. Fig.~\ref{Figure_1}(e) shows PL spectra for a bare NW, a WSe$_2$ monolayer on a flat substrate, and a strain-coupled monolayer integrated with the NW. The bare NW exhibits PL emission in the wavelength range of 825-850\,nm, corresponding to band-to-band electron-hole recombination in wurtzite InP \cite{dalacu2012ultraclean, laferriere2022unity}. The unstrained monolayer shows broad PL emission spanning 720-780\,nm with no distinct spectral features, indicating the absence of quantum emission. This observation is consistent with previous reports \cite{arora2015excitonic, huang2016probing}. In contrast, the strained monolayer displays multiple narrow and bright peaks in its PL spectrum superimposed on the broad emission band. The emission observed in the 825-850\,nm range arises from the underlying NW.
\begin{figure}[]
\centering
\includegraphics[scale=0.26]{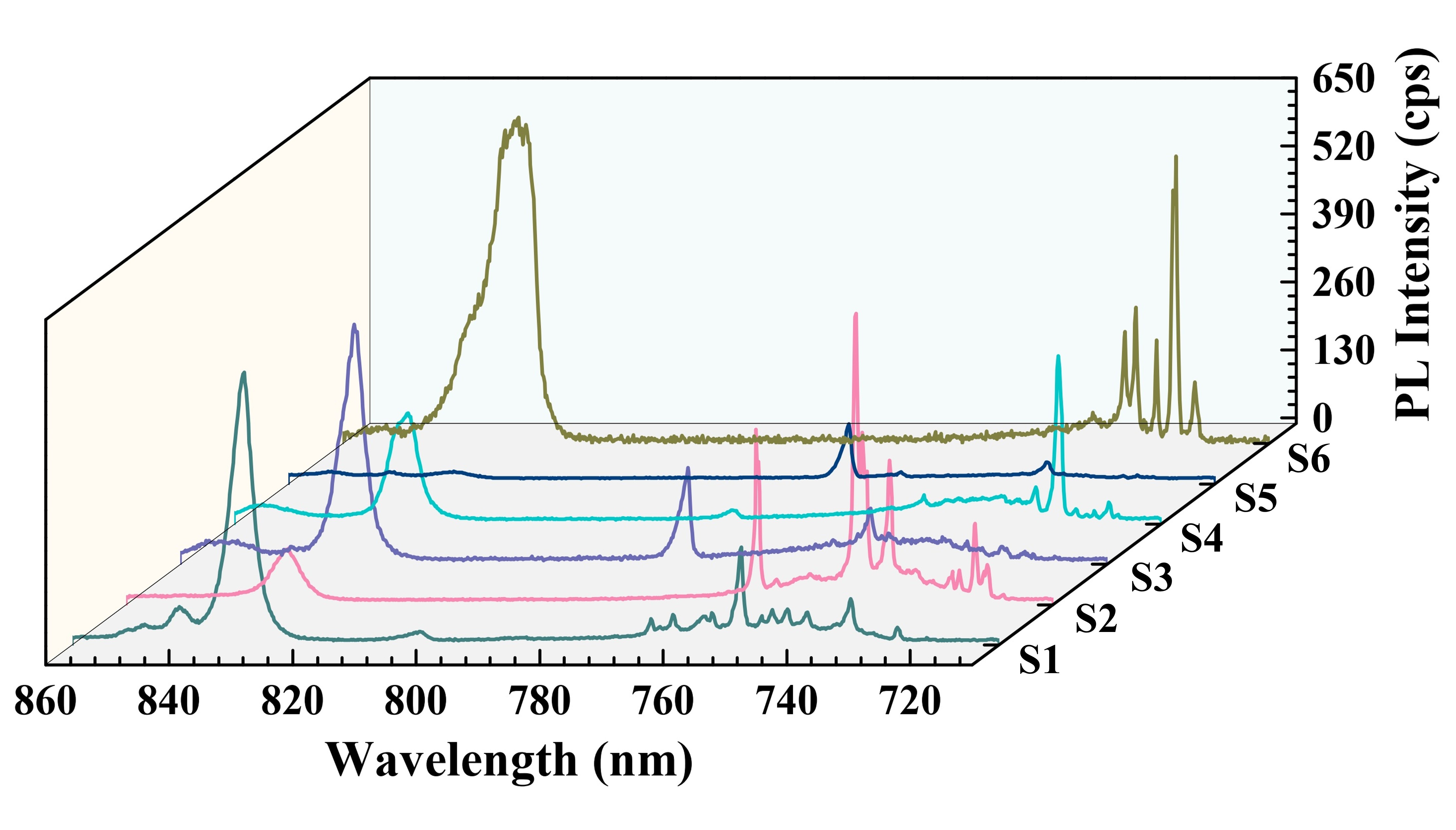}
\caption{PL spectra collected from different strained spots (S1 to S6) with excitation powers of 8\,$\mu$W (10\,$\mu$W for spot S6).}
\label{Figure_2}
\end{figure}
 Strain-induced emission sites were distributed throughout the sample and displayed bright and spectrally narrow PL lines, characteristic of localized strain-induced quantum emitters \cite{singh2025exploring, abramov2023photoluminescence, kumar2015strain, palacios2017large}. SEM images of several such locations are shown in Supplementary Figures A1 and A2. 
Fig.~\ref{Figure_2} presents the PL spectra of selected emission spots (S1-S6). PL spectra from other strained spots can be found in the supplementary information (see Supplementary Figure A3). Several narrow emission lines are observed at each position, indicating multiple localized emitters. Table \ref{table1} lists the most intense lines observed. The distinct emission profiles at each site suggest that the strain induced by the underlying NWs varies spatially. Despite their proximity on the same NW, spots S3 and S4 show markedly different PL spectra, indicating local variations in strain that influence their optical response. To further investigate the quantum dot-like emission characteristics, power-dependent PL measurements were performed. Fig.~\ref{Figure_3}(a-b) display the PL spectra of the strained monolayer at spots S2 and S3 under varying excitation powers (P).  The PL intensity, I(P), of the sharp emission peaks was fit using a saturation model described by the equation:
\begin{equation}
I(P) = I_{sat} \frac{P}{P + P_{sat}}
\end{equation}
where I$_{sat}$ is the saturation intensity and P$_{sat}$ is the saturation power. Fig.~\ref{Figure_3}(c-d) present the measured PL intensity for the emission line at 736.47\,nm and 757.70\,nm from the strained spot S2, yielding extracted P$_{sat}$ values of 12.4\,µW and 10.8\,µW, respectively. Fig.~\ref{Figure_3}(e) shows the measured PL intensity for the line at 778.37\,nm from strained spot S3, with an extracted P$_{sat}$ value of 3.4\,µW.

\begin{table}[h]
\resizebox{\columnwidth}{!}{%
\begin{tabular}{|c|c|c|c|c|}
\hline
Spot & Wavelength (nm)                                                           & Linewidth (nm)                                                  & $ g^{(2)} (0)$                                                              & $\tau_0$ (ns)                                                     \\ \hline
S1        & \begin{tabular}[c]{@{}c@{}}734.11\\ 751.91\end{tabular}                   & \begin{tabular}[c]{@{}c@{}}1.02\\ 1.11\end{tabular}               & -                                                               & -                                                           \\ \hline
S2        & \begin{tabular}[c]{@{}c@{}}736.49\\ 740.72\\ 742.00\\ 757.93\end{tabular} & \begin{tabular}[c]{@{}c@{}}0.90\\ 1.27\\ 0.78\\ 0.90\end{tabular}   & -                                                               & -                                                           \\ \hline
S3        & 778.37                                                                    & 2.11                                                              & 0.049$\pm$0.022                                                     & 5.38$\pm$0.18                                                   \\ \hline
S4        & 726.62                                                                    & 1.10                                                               & -                                                               & -                                                           \\ \hline
S5        & 769.69                                                                    & 1.90                                                               & 0.20$\pm$0.021                                                      & 4.62$\pm$0.17                                                   \\ \hline
S6        & \begin{tabular}[c]{@{}c@{}}726.70\\ 728.70\\ 731.65\\ 733.42\end{tabular} & \begin{tabular}[c]{@{}c@{}}1.03\\ 0.41\\ 0.67\\ 0.64\end{tabular} & \begin{tabular}[c]{@{}c@{}}0.11$\pm$0.057\\ 0.24$\pm$0.046\end{tabular} & \begin{tabular}[c]{@{}c@{}}1.9$\pm$0.17\\ 2.3$\pm$0.20\end{tabular} \\ \hline
\end{tabular}
}
\caption{Summary of measured emission characteristics across different spots.}
\label{table1}
\end{table}

\begin{figure}[h]
\centering
\includegraphics[scale=0.35]{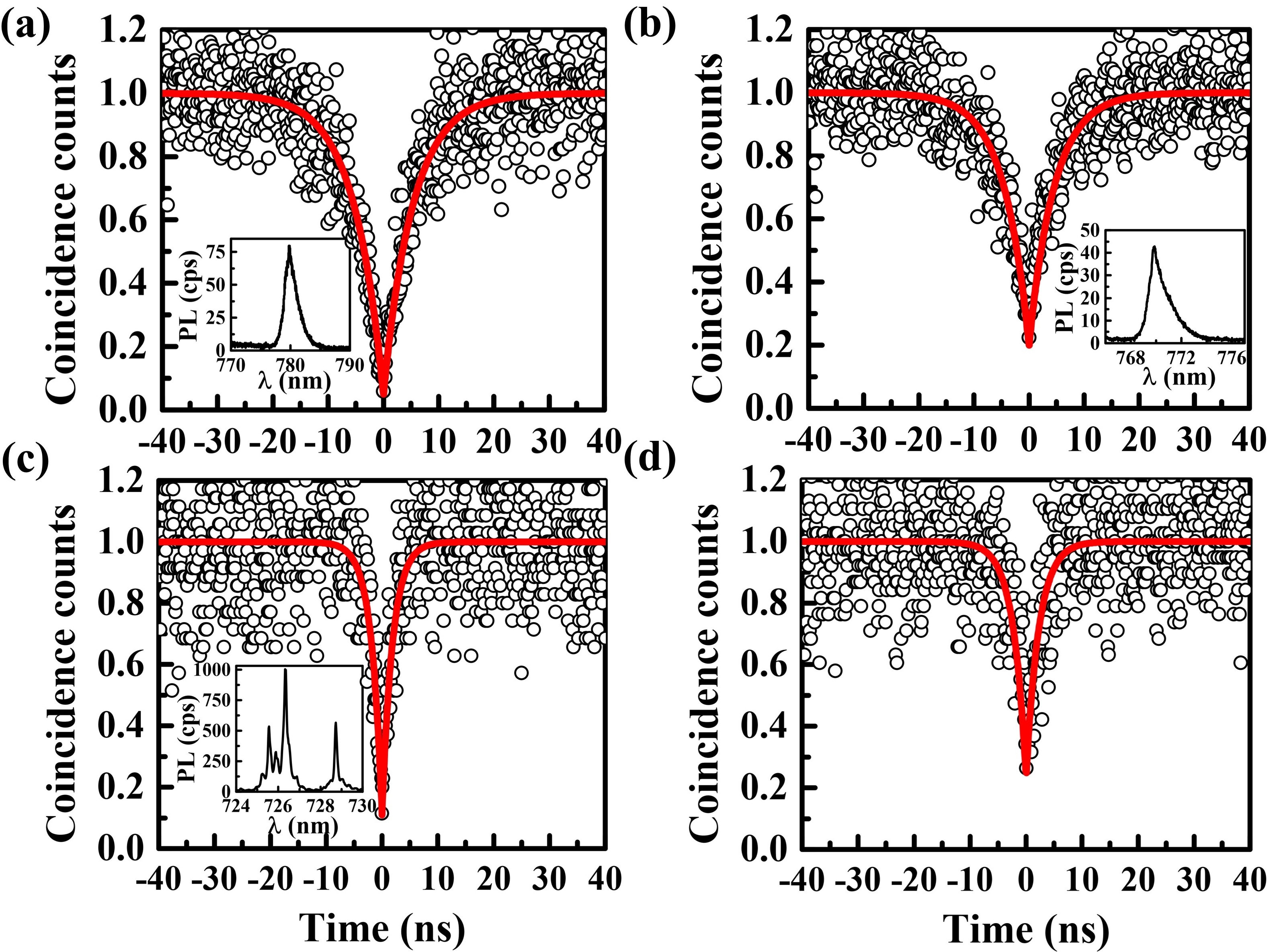}
\caption{Second-order autocorrelation measurement: (a) Quantum emission at 778.37\,nm from the strained spot S3 with an excitation power of 5\,$\mu$W. The inset shows the corresponding emission spectrum at an excitation power of 8~$\mu$W. (b) Quantum emission at 769.69\,nm from the strained spot S5 with an excitation power of 5~$\mu$W, with the inset displaying the respective spectrum at 8~$\mu$W. (c-d) Quantum emissions at 726.7~nm and 728.7~nm from the strained spot S6, both measured at an excitation power of 10~µW. The inset shows the corresponding emission spectrum at an excitation power of 10\,$\mu$W. Open circles represent raw data, while red solid lines indicate the fitted curves. PL spectra shown in the insets of (a)-(c) are collected using a 1200 g/mm grating.
}
\label{Figure_4}
\end{figure}

The brightest emission peaks were further characterized using second-order autocorrelation measurements.  The results of these experiments are shown in Fig.~\ref{Figure_4} for the strained spots S3, S5, and S6.  The experimental data were fitted using the relation:
\begin{equation}
  g^{(2)} (\tau)=1-Ae^{-|{\tau}/{\tau_0}|}   
\end{equation}
where A is a fitting parameter and $\tau_0$ is the characteristic decay time. The extracted values of $ g^{(2)} (0)$ and $\tau_0$ for the bright and narrow emission lines for various spots are listed in Table~\ref{table1}.
A range of values of g$^2$(0) are observed, extending as low as 0.049.
Similar quantum emission characteristics are observed across all strained monolayer regions, underscoring the robustness of the strain profile generated by the NWs in creating QEs. 

Strain plays a crucial role in enabling quantum emission in 2D materials, and identifying the optimal strain is essential to realize efficient single photon sources \cite{branny2017deterministic, palacios2017large, kumar2015strain, kim2022high, abramov2023photoluminescence, stevens2022enhancing}.  The wurtzite NWs have hexagonal cross-sections (see inset, Fig.~\ref{Figure_1}(b)) with sidewalls corresponding to two competing crystal facet orientations: \{11$\overline{2}$0\} or \{1$\overline{1}$00\} \cite{dalacu2009selective, dalacu2021tailoring, leitsmann2007surface}. Depending on the growth conditions, various scenarios can occur: both facet orientations may coexist, blurring the termination edges; one facet orientation may gradually transition into the other along the NW length; or the facets may merge towards the tapered tip (see inset of Fig.~\ref{Figure_1}(b) and Supplementary Figure A4). This facet-dependent geometry leads to a spatially varying strain distribution along the NWs which is further dependent on how the hexagonal NW is oriented with respect to the SiO$_2$/Si substrate. To quantify the local strain in the WSe$ _2 $ monolayer, we employ a model based on continuum elastic theory \cite{landau1970theory}, which relates out-of-plane deformation to in-plane tensile strain. Using AFM-derived height profiles, this method provides a non-destructive way to spatially map strain across the monolayer.  The out-of-plane strain ($\epsilon_{zz}$) is calculated using the expression:
\begin{equation}
\left| \epsilon_{zz}  \right|=\left| \frac{\nu d}{1-\nu}\left[ \frac{\partial^2 h}{\partial x^2} + \frac{\partial^2 h}{\partial y^2}\right] \right|
\end{equation}
Here, $\nu$ is the Poisson's ratio which is $\sim$ 0.2 for monolayer WSe$ _2 $ \cite{kang2013band, zeng2015electronic}, $d$ (= 0.8\,nm) is the thickness of a single monolayer of WSe$ _2 $, and $h(x,y)$ is the displacement out of the plane, which can be obtained from AFM. This formula sums the strain contributions from the curvature in the principal in-plane directions. This model assumes small deformations (significantly larger than the thickness) and linear elastic behavior, with strain originating purely from geometric curvature.  When a flexible monolayer conforms to a non-planar substrate, such as wrapping over NWs or nanopillars, it undergoes in-plane stretching even without external forces. This strain arises because the actual surface area of the deformed membrane becomes larger than its projected in-plane area, requiring the material to stretch to accommodate the curvature. The squared gradients in the formula quantify the steepness with which the membrane bends in each direction, directly relating out-of-plane deformation to in-plane tensile strain. 

\begin{figure*}
 \centering
\includegraphics[scale=0.5]{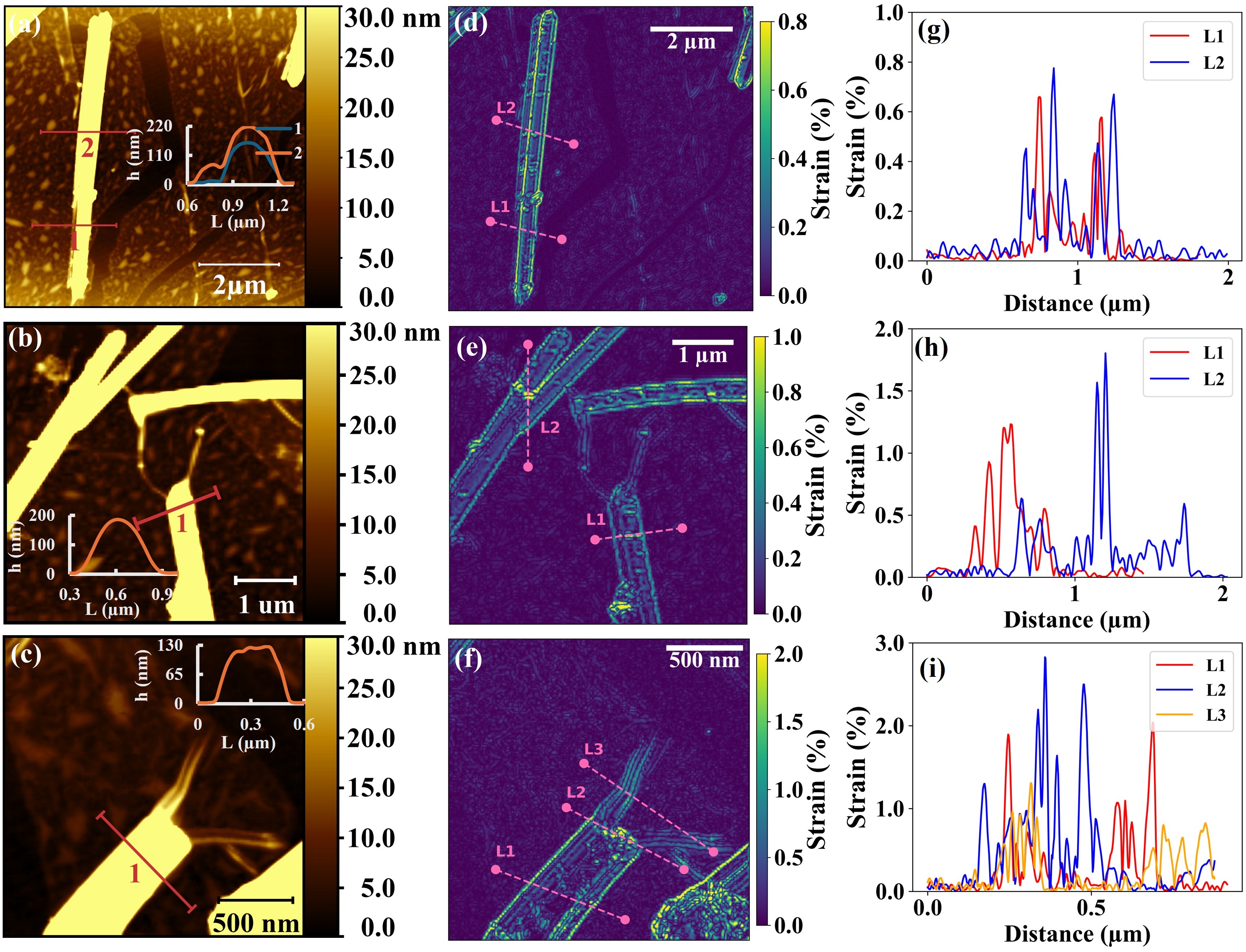}
\caption{Strain mapping of monolayer WSe$ _2 $ on NWs at location S1, S2, and S3. (a–c) AFM images with height profiles (insets) along the marked line cuts. (d–f) Strain maps computed from AFM data using continuum elastic theory. Line cuts indicate directions for strain extraction. (g–i) Comparison of localized strain profiles along the line cuts from (d-f).}
\label{Figure_5}
\end{figure*}

Fig.~\ref{Figure_5} presents a comprehensive strain analysis based on AFM images acquired from three distinct regions of the monolayer WSe$ _2 $ sample, labeled S1, S2 and S3. These labels are used consistently in the corresponding PL spectra, enabling a direct correlation between the local strain features and the associated PL emission.  Fig.~\ref{Figure_5}(a-c) show the AFM topographies for regions S1, S2, and S3, respectively. The red linecuts in each figure correspond to an area where quantum emission from the monolayer is observed. The height profile along these linecuts is presented in the inset of each figure, highlighting localized topographical deformations induced by the underlying InP NWs. Fig.~\ref{Figure_5}(d–f) show the corresponding out-of-plane strain maps calculated with continuum elastic theory using the AFM data. Strain line profiles shown in Fig.~\ref{Figure_5}(g–i), extracted along the marked paths (L1, L2, etc.) in Fig.~\ref{Figure_5}(d–f), respectively, highlight the localized and direction-dependent nature of the strain. These maps reveal non-uniform and anisotropic strain fields, primarily concentrated around NW-induced topographical features. 

Fig.~\ref{Figure_5}(a) displays a long NW ($\sim$7~$\mu$m in length) with asymmetric end diameters of $ \sim $250\,nm and $ \sim $150\,nm lying fully beneath the monolayer. The two linecuts capture this tapering geometry and reveal the resulting asymmetry in monolayer deformation, which arises from the gradual blurring of distinct facets in the tapered region. The maximum strain occurs at the edges, reaching $ \sim $0.8\%. In contrast, in Fig.~\ref{Figure_5}(b), the line cut profile indicates that the top surface of the NW corresponds to the edge formed by two adjacent facets, resulting in an out-of-plane displacement of approximately 200\,nm.  As shown in Fig.~\ref{Figure_5}(h), this leads to the largest strain at the top/center of the NW.   The AFM profile in Fig.~\ref{Figure_5}(c) reveals a flat NW facet oriented upward, inducing strain in the monolayer along the two edges where it drapes down to the substrate. This leads to a maximum strain of approximately 2\% at the edges of the NW, as shown for the line cut L1 in Fig.~\ref{Figure_5}(i).  The line cuts L2 and L3 show that strain values as high as $ \sim $2.9\% also appears at the ends of the NW and within the wrinkled, tent-like region away from the NW.  These findings demonstrate that nanoscale geometric features such as NW height, width, edge sharpness, and facet orientation relative to the substrate play crucial role in shaping the spatial distribution and magnitude of strain in 2D materials.

Our previous work on bilayer WSe$ _2 $ established that nanopillars with larger height-to-width aspect ratios produce more localized and intense strain, resulting in the formation of well-defined quantum emitters \cite{singh2025exploring}. Enhanced PL intensity and spectral purity were observed in regions corresponding to the smallest pillar diameters and highest aspect ratios. In the present study, using monolayer WSe$ _2 $ positioned on InP NWs, we further identify the sharpness of the edge, the facet, and the contact geometry as critical factors governing the spatial distribution and magnitude of strain. Although smooth, gradually varying topographies lead to broader and less concentrated strain fields, sharp structural transitions effectively confine strain along well-defined regions.   Unlike nanopillars, the aspect ratio of NWs in this study does not play a dominant role in strain tuning, as the vertical displacement of the monolayer on NW is comparable to the diameter of the NW, resulting in an effective aspect ratio close to unity. Our results combined with prior studies demonstrate that optimizing both aspect ratio and nanoscale morphological features provides an effective approach for the deterministic engineering of strain-tunable quantum emitters in 2D TMDs \cite{singh2025exploring, branny2017deterministic, palacios2017large}. 

Our demonstration of strain-induced QE creation in a monolayer WSe$_2$/InP NW hybrid structure lays the foundation for integrating TMD sources with traditional III-V photonics. Coupling of TMD QEs into NW waveguides containing InAsP quantum dots would enable the propagation of single photons with multiple emission wavelengths down the same waveguide.  Our experiments indicate that multiple QE sites may be created along a single NW.  These would permit multimode single photon routing needed for multiplexing of quantum photonic channels. With suitable selection of emission wavelengths, cascaded absorption may also enable transduction between guided and surface emission geometries. For such applications, the faceted NW surfaces provide flexibility in tailoring the strain-induced QE locations relative to the coupled modes of the waveguide.  While TMD QEs induced using rectangular waveguide geometries have been shown to result in QEs primarily at the edges with waveguide coupling efficiency $<$1\% \cite{ErrandoHerranz:2021,blauth2018coupling}, a ten-fold increase in coupling efficiency is expected for a QE oriented at the top of the waveguide \cite{ErrandoHerranz:2021}, a value that could be further enhanced through the use of optical cavities \cite{peyskens2019integration}.  For NW waveguide geometries such as in Fig.~\ref{Figure_5}(b), in which an apex is formed at the top corresponding to the interface of two side facets, QEs are expected at the top surface and may lead to better waveguide coupling. Together with recent efforts to integrate InP NWs containing semiconductor quantum dots with SiN waveguides \cite{Yeung:2023} and the realization of on-chip superconducting NW single photon detectors \cite{Tanner:2012}, our findings will support efforts to develop scalable, fully-integrated quantum photonic systems.

\section{Conclusion}
In summary, we have demonstrated the creation of QEs in WSe$_2$ through transfer of single monolayers onto InP NWs.  The conformation of the monolayer to the faceted NW surface leads to anisotropic strain, providing flexible control of QE position and optical properties.  Characterization of the level of strain using AFM and continuum elastic theory shows that NW facet edges lead to a strain that varies from 0.3 to 2.9 \%.  These strain features give rise to bright and spectrally narrow emission peaks, with linewidths ranging from 0.4 to 2 nm and g$^2$(0) as low as 0.049. Our experiments highlight the potential for integrating traditional III-V QEs with strain-induced TMD emitters into a single platform, laying the groundwork for scalable architectures for quantum communication and information processing.

\begin{acknowledgments}
This research was supported by the Natural Sciences and Engineering Research Council of Canada, Grant No. RGPIN-2020-06322, and the Internet of Things: Quantum Sensors Challenge Program at the National Research Council of Canada.
\end{acknowledgments}

\section*{Data Availability Statement}
The data that support the findings of this study are available from the corresponding author upon reasonable request.

\bibliographystyle{aipnum4-1}
\bibliography{ref}

\end{document}